\documentclass[fleqn,10pt]{SelfArx} 

\definecolor{color1}{RGB}{0,0,90} 
\definecolor{color2}{RGB}{0,20,20} 

\usepackage{hyperref} 
\hypersetup{hidelinks,colorlinks,breaklinks=true,urlcolor=color2,citecolor=color1,linkcolor=color1,bookmarksopen=false,pdftitle={Title},pdfauthor={Author},urlcolor=blue}

\usepackage{graphicx}
\usepackage[square]{natbib}
\usepackage{amsmath}
\usepackage{multirow}
\usepackage{subfigure}
\usepackage{pdflscape}

\newcommand{\pd}{\partial}
\newcommand{\n}[1]{\mathrm{#1}}
\newcommand{\LaFeCoSi}{LaFeCoSi}

\JournalInfo{Published in Journal of Magnetism and Magnetic Materials, Vol. 322 (24), 3882–3888, 2010} 
\Archive{\href{http://dx.doi.org/10.1016/j.jmmm.2010.08.013}{DOI: 10.1016/j.jmmm.2010.08.013}} 

\PaperTitle{Magnetocaloric properties of LaFe$_{13-x-y}$Co$_x$Si$_y$ and commercial grade Gd} 

\Authors{R. Bj\o{}rk$^1$, C. R. H. Bahl$^1$ and M. Katter$^2$} 
\affiliation{\textsuperscript{1}\textit{Department of Energy Conversion and Storage, Technical University of Denmark - DTU, Frederiksborgvej 399, DK-4000 Roskilde, Denmark}} 
\affiliation{\textsuperscript{2}\textit{Vacuumschmelze GmbH \& Co. KG, D-63450 Hanau, Germany}} 
\affiliation{*\textbf{Corresponding author}: rabj@dtu.dk} 

\newcommand{\keywordname}{Keywords} 

\Abstract{The magnetocaloric properties of three samples of LaFe$_{13-x-y}$Co$_x$Si$_y$ have been measured and compared to measurements of commercial grade Gd. The samples have (x=0.86, y=1.08), (x=0.94, y=1.01) and (x=0.97, y=1.07) yielding Curie temperatures in the range 276-288 K. The magnetization, specific heat capacity and adiabatic temperature change have been measured over a broad temperature interval. Importantly, all measurements were corrected for demagnetization, allowing the data to be directly compared. In an internal field of 1 T the maximum specific entropy changes were 6.2, 5.1 and 5.0 J/kg K, the specific heat capacities were 910, 840 and 835 J/kg K and the adiabatic temperature changes were 2.3, 2.1 and 2.1 K for the three \LaFeCoSi{} samples respectively. For Gd in an internal field of 1 T the maximum specific entropy change was 3.1 J/kg K, the specific heat capacity was 340 J/kg K and the adiabatic temperature change was 3.3 K. The adiabatic temperature change was also calculated from the measured values of the specific heat capacity and specific magnetization and compared to the directly measured values. In general an excellent agreement was seen.}

\begin{document}

\flushbottom 

\maketitle 


\thispagestyle{empty} 

\section{Introduction}
The magnetocaloric effect (MCE) is observed as a temperature change of a magnetic material when this is subjected to a changing external magnetic field. Such magnetocaloric materials (MCM) are interesting with respect to magnetic refrigeration, which is an emerging refrigeration technology based on the MCE that aims to provide environmentally friendly energy efficient cooling.

If an MCM with a positive magnetocaloric effect is subjected to a magnetic field and the conditions are kept adiabatic the temperature of the MCM will increase by the adiabatic temperature change, $\Delta{}T_\mathrm{ad}$. Had the conditions been kept isothermal the specific entropy would instead have been reduced by the isothermal entropy change, $\Delta{}s$. Both $\Delta{}T_\mathrm{ad}$ and $\Delta{}s$ are functions of temperature and magnetic field. These two properties, along with the specific heat capacity, $c_\n{p}$, which is also a function of temperature and magnetic field, are the three most important properties of an MCM with regard to application in magnetic refrigeration. Secondary properties such as the thermal conductivity, density and porosity can also be of importance, although these are in general not strong functions of temperature and magnetic field. A substantial number of magnetocaloric materials are known, each of whose properties has a different dependence on temperature and magnetic field \cite{Gschneidner_2005}. The MCE is generally largest near the phase transition of the MCM, known as the Curie temperature, $T_\n{C}$.

For an MCM with a second order phase transition the measured magnetocaloric data are related as the adiabatic temperature change can be calculated as
\begin{eqnarray}\label{Eq_dTad}
\Delta{}T_\mathrm{ad} = -\mu_0 \int_{H_\mathrm{i}}^{H_\mathrm{f}} \frac{T}{c_\mathrm{p}} \left(\frac{\pd m}{\pd T}\right)_{H}dH~,
\end{eqnarray}
once $c_\n{p}$ and $m$ are known.

For magnetic refrigeration an MCM must have a $T_\n{C}$ that is around room temperature. As can be seen from Eq. (\ref{Eq_dTad}) the change in magnetization must be substantial to provide a large $\Delta{}T_\mathrm{ad}$. Also, a high heat capacity provides a high thermal mass while a low heat capacity can cause a high adiabatic temperature change. The benchmark magnetocaloric material used in magnetic refrigeration is gadolinium (Gd), which has a $T_\mathrm{C}$ around 293 K and a $\Delta{}T_\mathrm{ad}$ of $\sim 3.5$ K at $T_\mathrm{C}$ in a field of 1 T \cite{Dankov_1998}. However, the Curie temperature of Gd cannot be tuned, and so the adiabatic temperature change will be low if the magnetic refrigerator is operated far from the Curie point.

In this paper we consider the properties of the magnetocaloric material, LaFe$_{13-x-y}$Co$_x$Si$_y$, referred to as \LaFeCoSi{}, which displays a significant adiabatic temperature change and has a tuneable Curie temperature. The precursor of this material, LaFe$_{13-x}$Si$_x$, which has a NaZn$_{13}$-type lattice crystal structure, has a magnetocaloric effect due to an itinerant electron metamagnetic transition from a paramagnetic to a ferromagnetic state \cite{Fujita_1999}.

Previous measurements of the specific magnetization and the specific entropy change of different LaFe$_{13-x-y}$Co$_x$Si$_y$ compounds have been reported \cite{Hu_2002a,Hu_2002b,Liu_2003,Passamani_2007,Saito_2007,Katter_2008,Yan_2008} as well as direct measurements of the adiabatic temperature change \cite{Hu_2005, Ilyn_2005, Balli_2009}, and a single measurement of the specific heat capacity \cite{Hansen_2009}. However, the published data are usually very widely spaced in external magnetic field and in general the published data have not been corrected for demagnetization effects, thus not allowing the different data to be compared. Also for \LaFeCoSi{} no comparison between a measured adiabatic temperature change and a temperature change calculated using Eq. (\ref{Eq_dTad}) have been reported. Importantly, if this relation can be verified experimentally only magnetization and heat capacity need be measured to fully characterize the magnetocaloric properties of a magnetocaloric material.

Here we have measured each of the magnetocaloric properties, i.e. the magnetization, specific heat capacity and adiabatic temperature change, of three different sintered samples of LaFe$_{13-x-y}$Co$_x$Si$_y$. The chemical composition of these are LaFe$_{11.06}$Co$_{0.86}$Si$_{1.08}$, LaFe$_{11.05}$Co$_{0.94}$Si$_{1.01}$ and \\ LaFe$_{10.96}$Co$_{0.97}$Si$_{1.07}$. These samples will be referred as Samples 1, 2 and 3 respectively.

The properties of the \LaFeCoSi{} samples are compared with the properties of commercial grade gadolinium, here simply termed Gd. This gadolinium is much cheaper than pure gadolinium, but the purity is also lower. The commercial grade gadolinium contains 99.5\% rare earth metal, of which 99.94\% are gadolinium. This Gd grade has previously been used in an actual magnetic refrigeration device \cite{Bahl_2008}.

The properties of both pure and impure gadolinium have previously been analyzed, and the main conclusions of this analysis were that the impure gadolinium has a lower adiabatic temperature change and, depending on the amount and types of impurities of the sample, a small shift of a couple of degree of the Curie temperature is also seen \cite{Dankov_1998}.

\section{Experimental setup}
The magnetocaloric properties of both \LaFeCoSi{} and commercial grade Gd have been measured using different lab equipment at Ris\o{} DTU.

Raw blocks of \LaFeCoSi{} were prepared by powder metallurgy as described by Ref. \cite{Katter_2008}. The sintered blocks were cut into plates with dimensions $25\times{}20\times{}0.9$ mm$^3$ utilizing the thermally induced decomposition and recombination (TDR) process \cite{Katter_2009}. From these plates, the samples for the various measurements were cut. The commercial grade Gd was obtained from China Rare Metal Material Co. in plates with dimensions $25\times{}40\times{}0.9$ mm$^3$.

The magnetization was measured using a LakeShore 7407 Vibrating Sample Magnetometer (VSM). Isothermal magnetization measurements as a function of field were made at a ramp rate of 2.5 mT/s up to a maximum field of $\mu_0 H_\mathrm{ext} = 1.6$ T and data were measured for every 5 mT. In a sample interval of $\pm 10$ K around the Curie temperature the measurements were taken at 1 K separation, while further from $T_\n{C}$ the separation was larger. Data were measured from 250 K to 310 K.

Once the magnetization has been determined the change in specific entropy can be calculated using
\begin{eqnarray}\label{Eq.dsm}
\Delta{}s_\mathrm{M} = \mu_0 \int_{H_\mathrm{i}}^{H_\mathrm{f}}\left(\frac{\pd m}{\pd T}\right)_{H}dH~,
\end{eqnarray}
where $H_\mathrm{i}$ is the initial magnetic field and $H_\mathrm{f}$ is the final magnetic field.

Calorimetric data in applied fields up to $\mu_0 H_\mathrm{ext} =1.4$ T were obtained using a differential scanning calorimeter (DSC) with the magnetic field provided by a concentric Halbach cylinder, which is an adjustable permanent magnetic field source \cite{Jeppesen_2008}. The specific heat capacity was measured at external field values of $\mu_0H_\mathrm{ext}=0,\;0.25,\;0.50,\;0.75,\;1.00\; \n{and}\; 1.40$ T in a temperature interval from 250 K to 310 K with a ramp of 1 K/min and from 310 K to 250 K with a ramp of -1 K/min. If no hysteresis is observed the calculated value of $c_\mathrm{p}$ from the two data sets are averaged. Data was taken every 12.5 ms. The DSC was calibrated using both copper and titanium reference samples.

Finally, the adiabatic temperature change was measured using an instrument designed at Ris\o{} DTU. A pneumatic piston  moves a sample holder in and out of a magnetic field generated by the same permanent magnet as used for the DSC. A sliding track ensures that the direction of the sample holder with respect to the magnetic field is fixed. The entire setup is placed in a freezer, which is used to control the temperature. The ramp rate of the temperature was controlled by a 75 W light bulb and heat leakage to the surroundings. In general a complete sweep in temperature took around 9 hours, during which time data were recorded. Due to this substantial time span, parasitic temperature gradients were kept as small as possible. The pneumatic piston was moved in and out of field every 5 seconds, with an actual movement time of no more than 100 ms. Because of the high sweep rate and the thermal isolation conditions can be assumed to be close to adiabatic. The temperature of the sample was recorded every 100 ms.

The sample is prepared by placing a type E thermocouple between two equally sized plates of the given sample which are then glued together. The sample is packed in isolating foam and placed in a sample holder together with a Hall probe (AlphaLab Inc, Model: DCM). Data were measured at external field values of $\mu_0H_\mathrm{ext}=0.25,\;0.50,\;0.75,\;1.00\; \n{and}\; 1.40$ T in a temperature interval from 250 K to 310 K. For each of the different materials all measurements were done on samples cut from the same large sample plate. The magnet is also cooled in the setup. This changes the magnetic field produced by the magnet slightly, but at 1 T the change is less than 3\%.

\begin{figure*}[!t]
\centering
\subfigure[Specific entropy change of Sample 1.]{\includegraphics[width=0.47\textwidth]{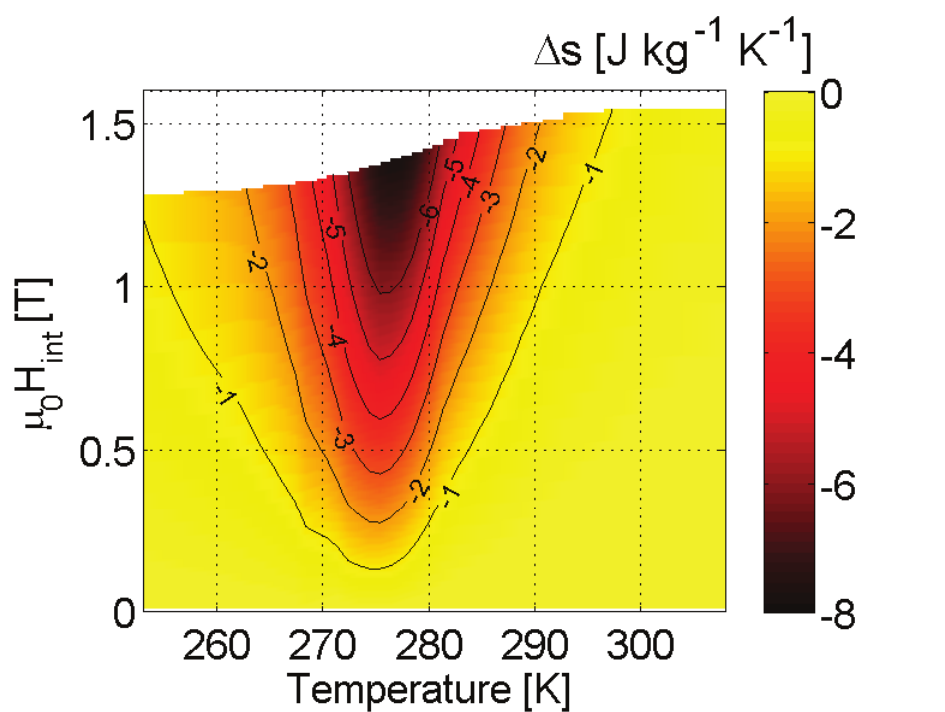}}\hspace{0.2cm}
\subfigure[Specific entropy change of Sample 2.]{\includegraphics[width=0.47\textwidth]{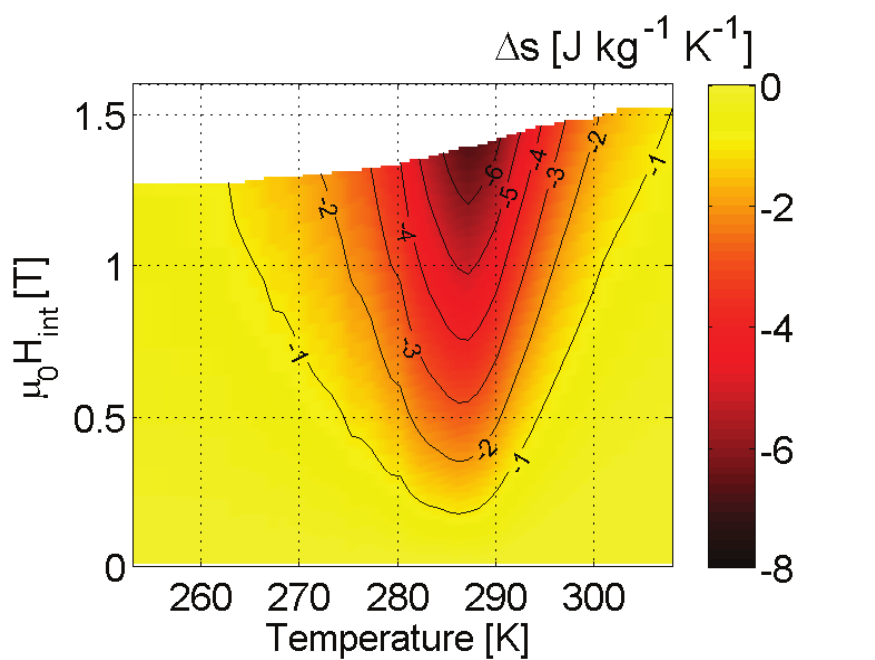}}\\
\subfigure[Specific entropy change of Sample 3.]{\includegraphics[width=0.47\textwidth]{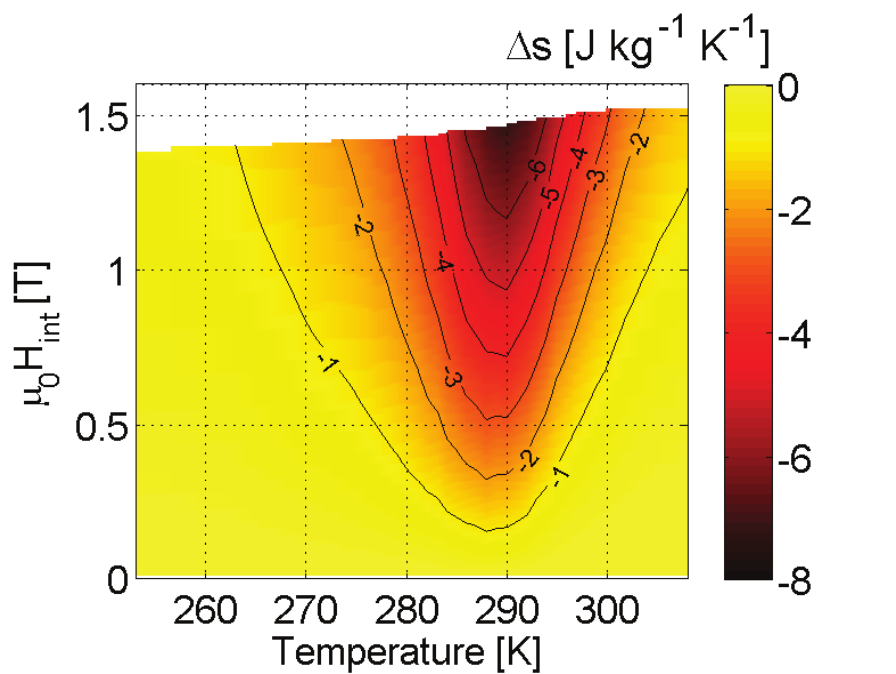}}\hspace{0.2cm}
\subfigure[Specific entropy change of Gd]{\includegraphics[width=0.47\textwidth]{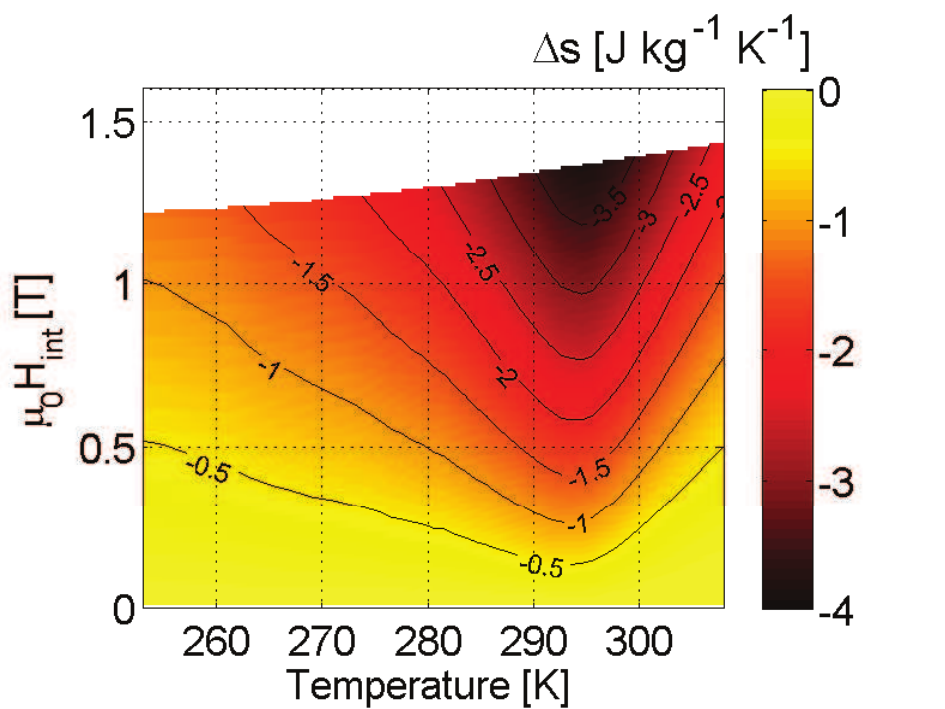}}
\caption{(Color online) The change in specific entropy, $\Delta{}s$, as calculated from Eq. (\ref{Eq.dsm}) as a function of temperature and internal field.}\label{Fig.Deltas}
\end{figure*}

\subsection{Internal magnetic field}\label{Sec.Internal_magnetic_field}
When subjecting a magnetic sample to an external magnetic field the internal field within the sample will depend on the geometry of the sample. It is extremely important to always report magnetocaloric properties as a function of internal field, as comparison with other or even identical materials is otherwise not meaningful \cite{Bahl_2009}. If the sample has an ellipsoidal shape the internal field will be homogeneous and can be calculated if the magnetization of the sample is known \cite{Osborn_1945}. If the sample has a non ellipsoidal shape the internal field will not be homogeneous across the sample. However, an average internal magnetic field can be found. This average internal magnetic field, $H_\mathrm{int}$, can be found by subtracting the demagnetization field $H_\mathrm{d} = N_\mathrm{d}M$, where $N_\mathrm{d}$ is the average demagnetization factor and $M$ is the magnetization, from the applied external field, $H_\mathrm{ext}$,
\begin{eqnarray}
H_\mathrm{int} = H_\mathrm{ext} - N_\mathrm{d}M~.
\end{eqnarray}
Assuming that the magnetic anisotropy is negligible the internal field, the external field and the magnetization are all very close to being parallel, so that only the magnitudes need be considered.

If the sample has a rectangular (orthorhombic) shape the average demagnetization factor can be calculated according to Ref. \cite{Aharoni_1998}. This factor is a good approximation to the true demagnetization field \cite{Smith_2010}.

Using the dimensions of the individual sample pieces and their respective demagnetization factors, given in Table \ref{Table.LaFeCoSi_and_Gd_dimensions}, all measurements presented here have been corrected for demagnetization by the approximation in Ref. \cite{Aharoni_1998}. Thus, all measured properties are reported as a function of internal field. While this correction is easy to perform it is not commonly done. In the papers previously published on \LaFeCoSi{} corrections for demagnetization has not been performed making comparison with the present work difficult.

\begin{table*}[!t]
\begin{center}
\caption{The dimensions and demagnetization factors for the different samples of \LaFeCoSi{} and Gd. The direction of the magnetic field is always along the $c$-axis.}\label{Table.LaFeCoSi_and_Gd_dimensions}
\begin{tabular}{lc|ccccr}
 & & \multicolumn{3}{c}{LaFe$_{13-x-y}$Co$_x$Si$_y$} & Gd \\
 & & $\left( \begin{array}{ll} x=0.86\\ y=1.08 \end{array} \right)$ & $\left( \begin{array}{ll} x=0.94\\ y=1.01 \end{array} \right)$ & $\left( \begin{array}{ll} x=0.97\\ y=1.07 \end{array} \right)$ & $\begin{array}{cc} \mathrm{commercial}\\ \mathrm{grade} \end{array} $ &
Unit\\
& & Sample 1 & Sample 2 & Sample 3 \\ \hline
Density              &                       & 6980  & 7290  & 7160  & 7900  & [kg m$^{-3}$]\\&\\
\multirow{2}{*}{VSM} & $a\times{}b\times{}c$ & $0.82\times{}3.27\times{}1.81$ & $0.85\times{}3.38\times{}1.96$ & $0.87\times{}2.04\times{}3.38$ & $0.90\times{}4.60\times{}2.26$ & [mm$^3$]\\
                     & $N_\mathrm{d}$        & 0.27   & 0.27   & 0.16   & 0.26   & [-]\\&\\
\multirow{2}{*}{DSC} & $a\times{}b\times{}c$ & $1.94\times{}0.85\times{}3.66$ & $2.47\times{}0.83\times{}2.50$ & $2.21\times{}0.85\times{}3.17$ & $2.04\times{}0.89\times{}2.78$ & [mm$^3$]\\
                     & $N_\mathrm{d}$        & 0.14   & 0.20   & 0.17   & 0.19 & [-]\\&\\
\multirow{2}{*}{$\Delta{}T_\n{ad}$} & $a\times{}b\times{}c$ & $7.36\times{}2.08\times{}5.33$ & $5.60\times{}1.74\times{}7.70$ & $9.90\times{}2.09\times{}10.23$ & $5.42\times{}2.01\times{}9.48$ & [mm$^3$]\\
                                    & $N_\mathrm{d}$        & 0.24   & 0.15   & 0.15   & 0.14   & [-]
\end{tabular}
\end{center}
\end{table*}

\section{Results and discussion}

The change in specific entropy was calculated by Eq. (\ref{Eq.dsm}) using the measured magnetization data and employing a numerical integration scheme. These results are shown in Figure \ref{Fig.Deltas}.

\begin{figure*}[!p]
\centering
\subfigure[Specific heat capacity of Sample 1.]{\includegraphics[width=0.47\textwidth]{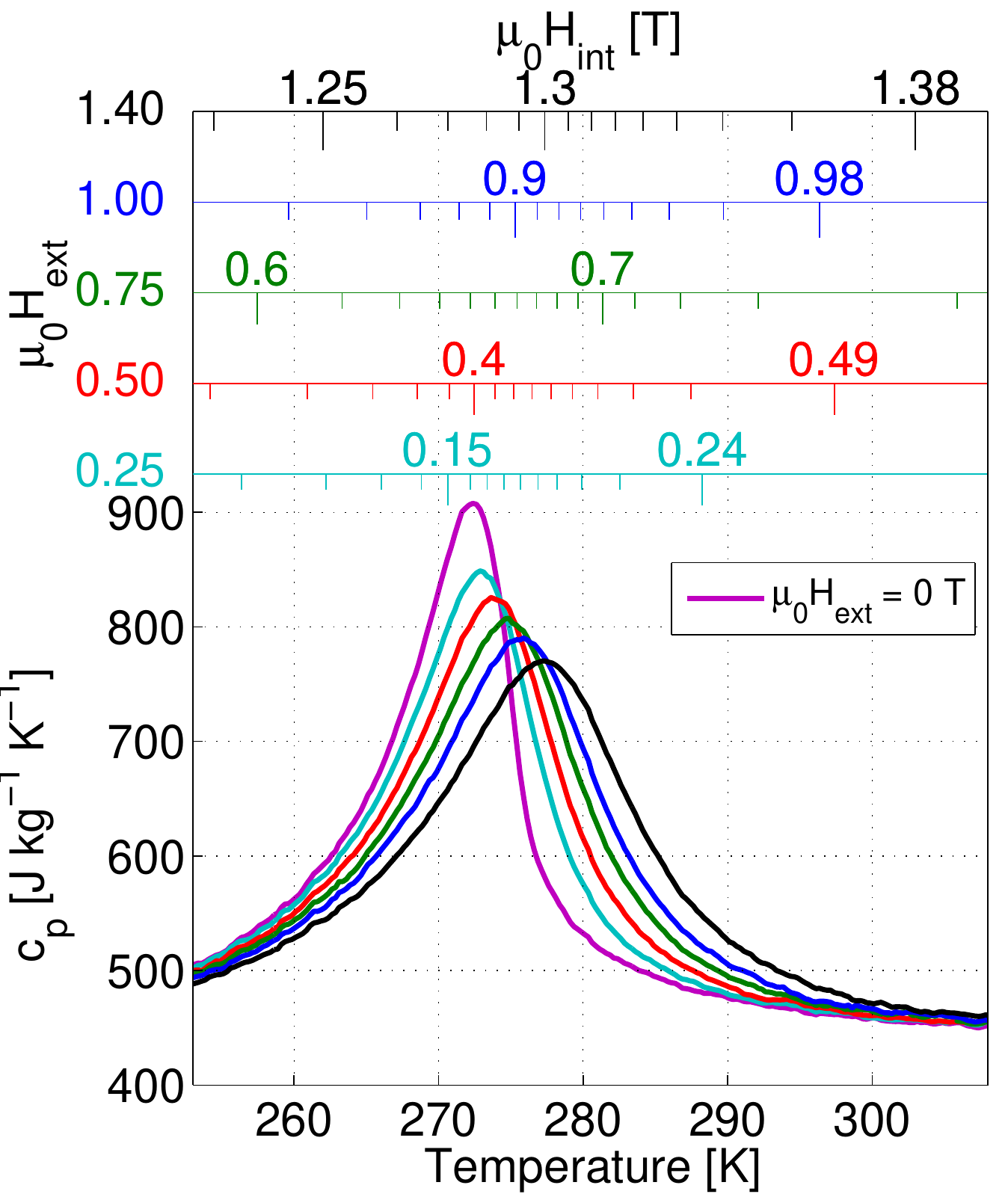}}\hspace{0.4cm}
\subfigure[Specific heat capacity of Sample 2.]{\includegraphics[width=0.47\textwidth]{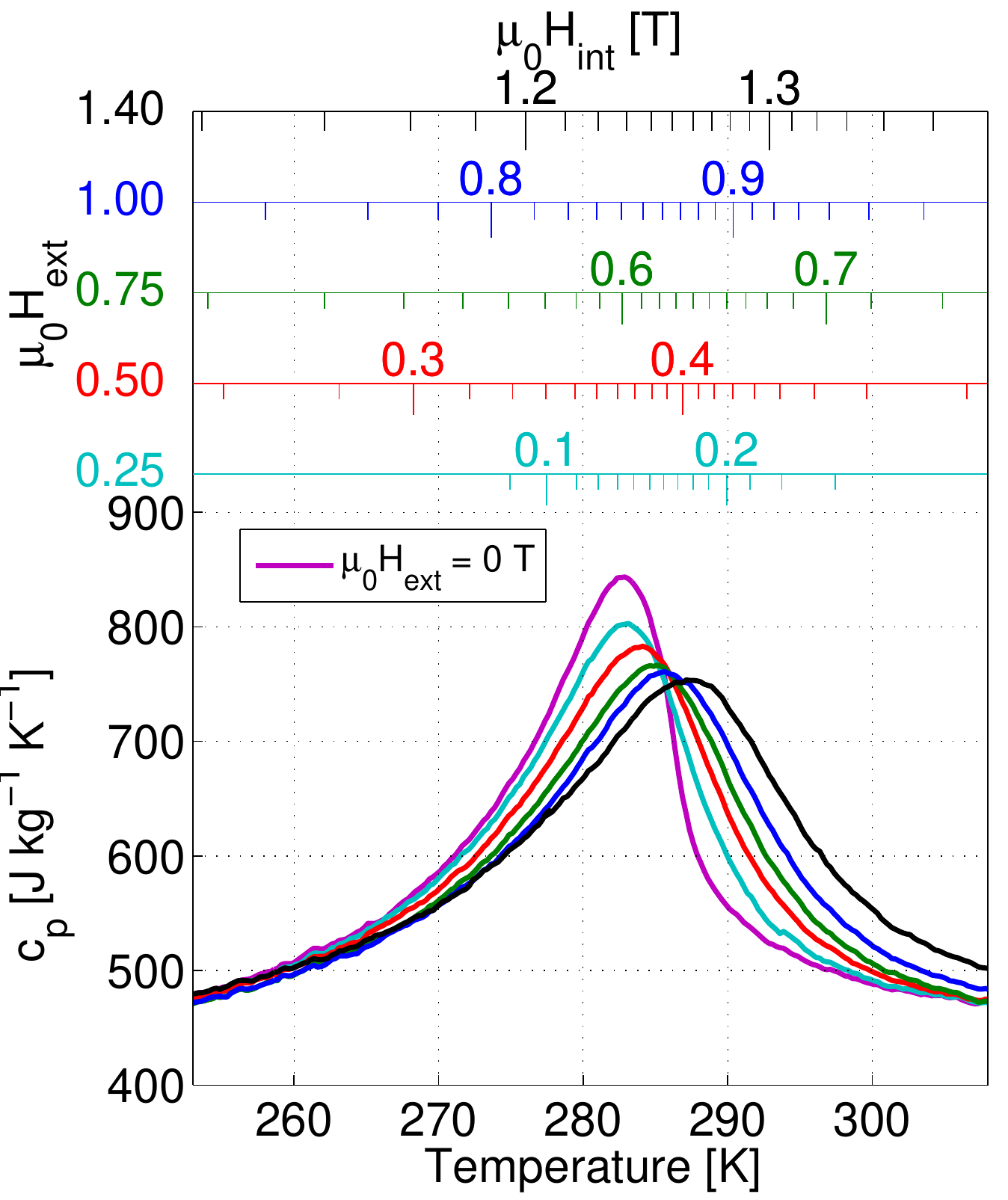}}\\
\subfigure[Specific heat capacity of Sample 3.]{\includegraphics[width=0.47\textwidth]{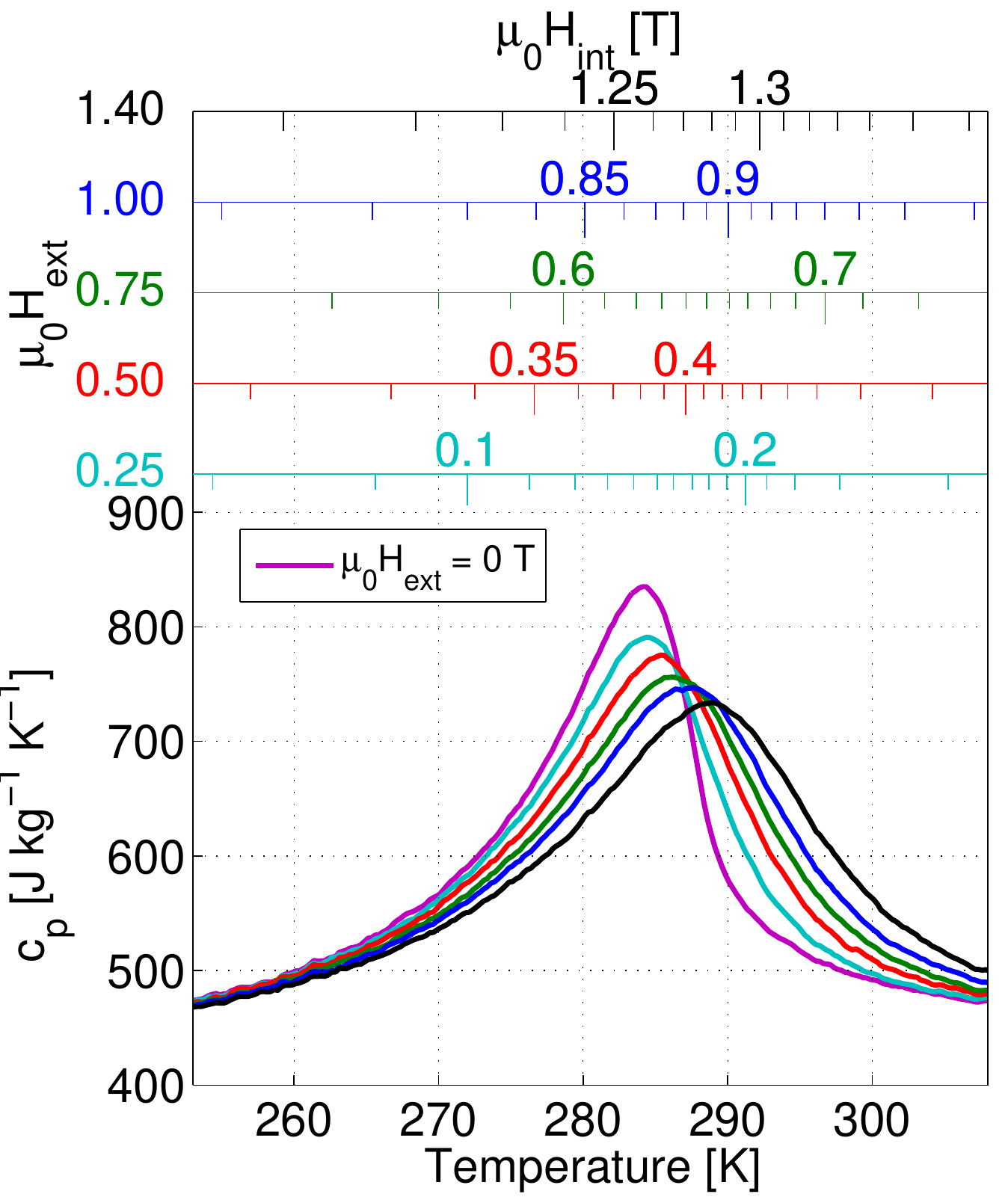}}\hspace{0.4cm}
\subfigure[Specific heat capacity of Gd]{\includegraphics[width=0.47\textwidth]{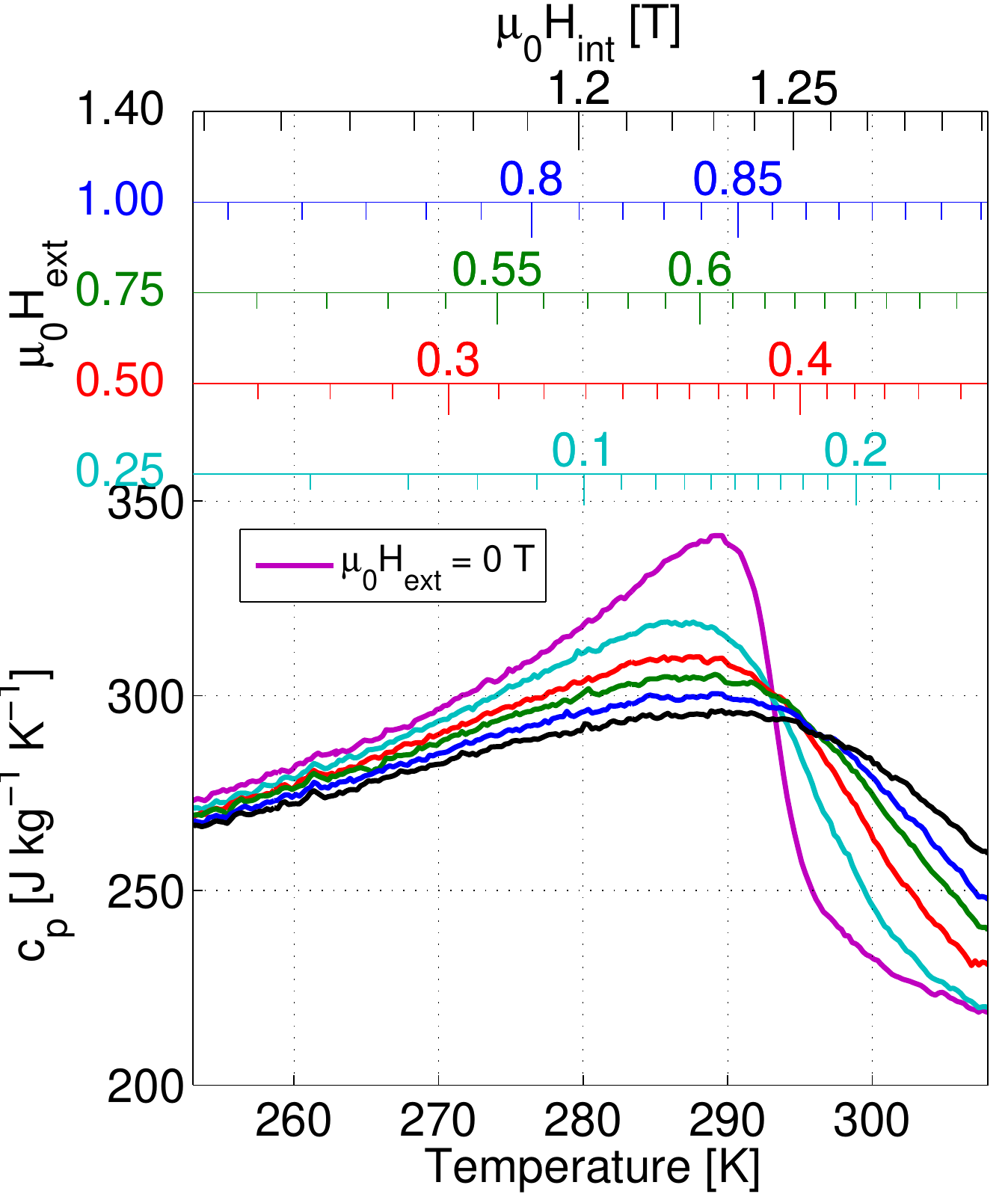}}
\caption{(Color online) The specific heat capacity, $c_\n{p}$, as a function of temperature for the internal field given in the top x-axis. Each of the top x-axis gives the internal field for the same colored curve. As the external field is increased the maximum value of the specific heat capacity decreases.}\label{Fig.cp}
\end{figure*}

\begin{figure*}[!p]
\centering
\subfigure[Adiabatic temperature change of Sample 1.]{\includegraphics[width=0.47\textwidth]{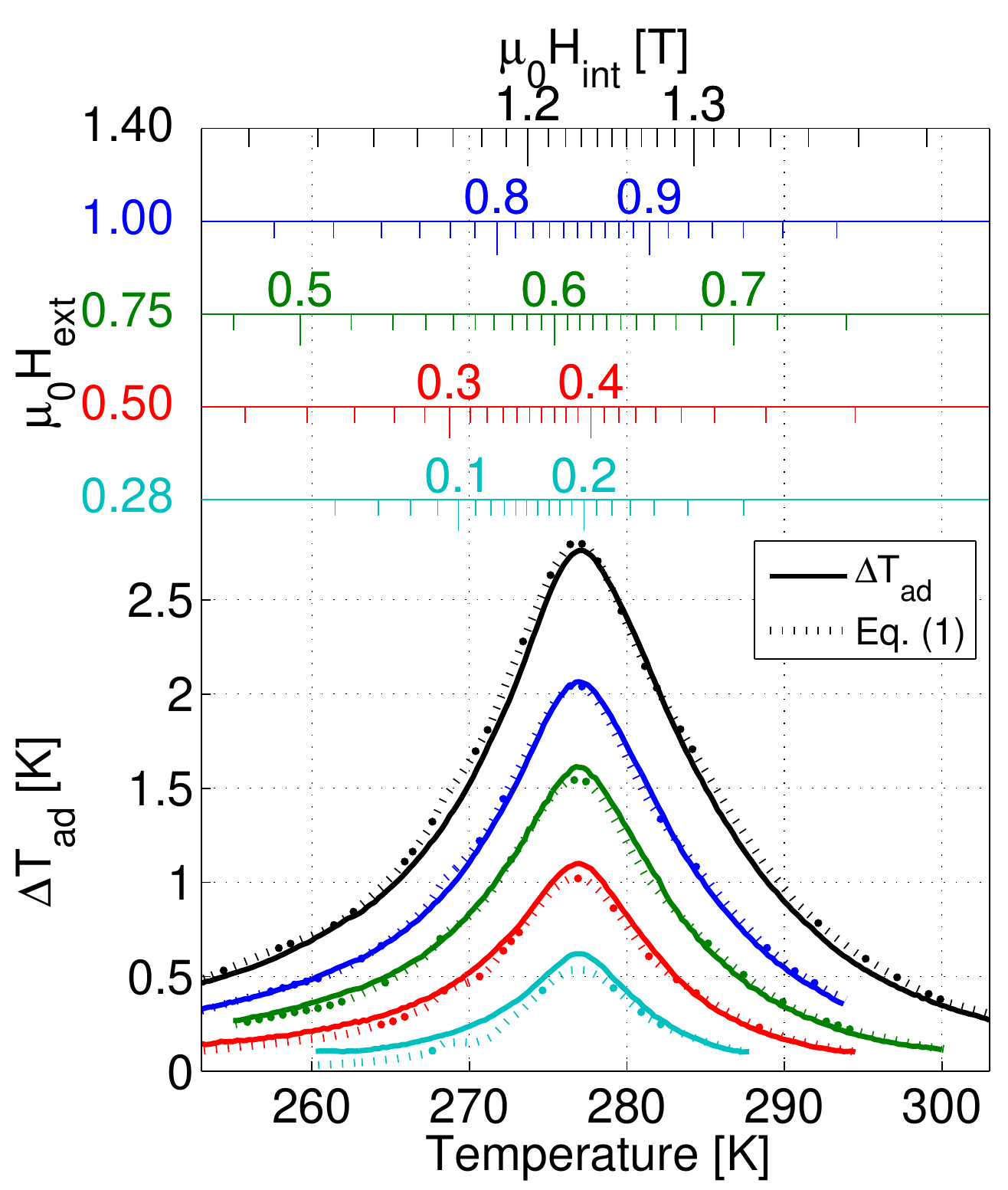}}\hspace{0.4cm}
\subfigure[Adiabatic temperature change of Sample 2.]{\includegraphics[width=0.47\textwidth]{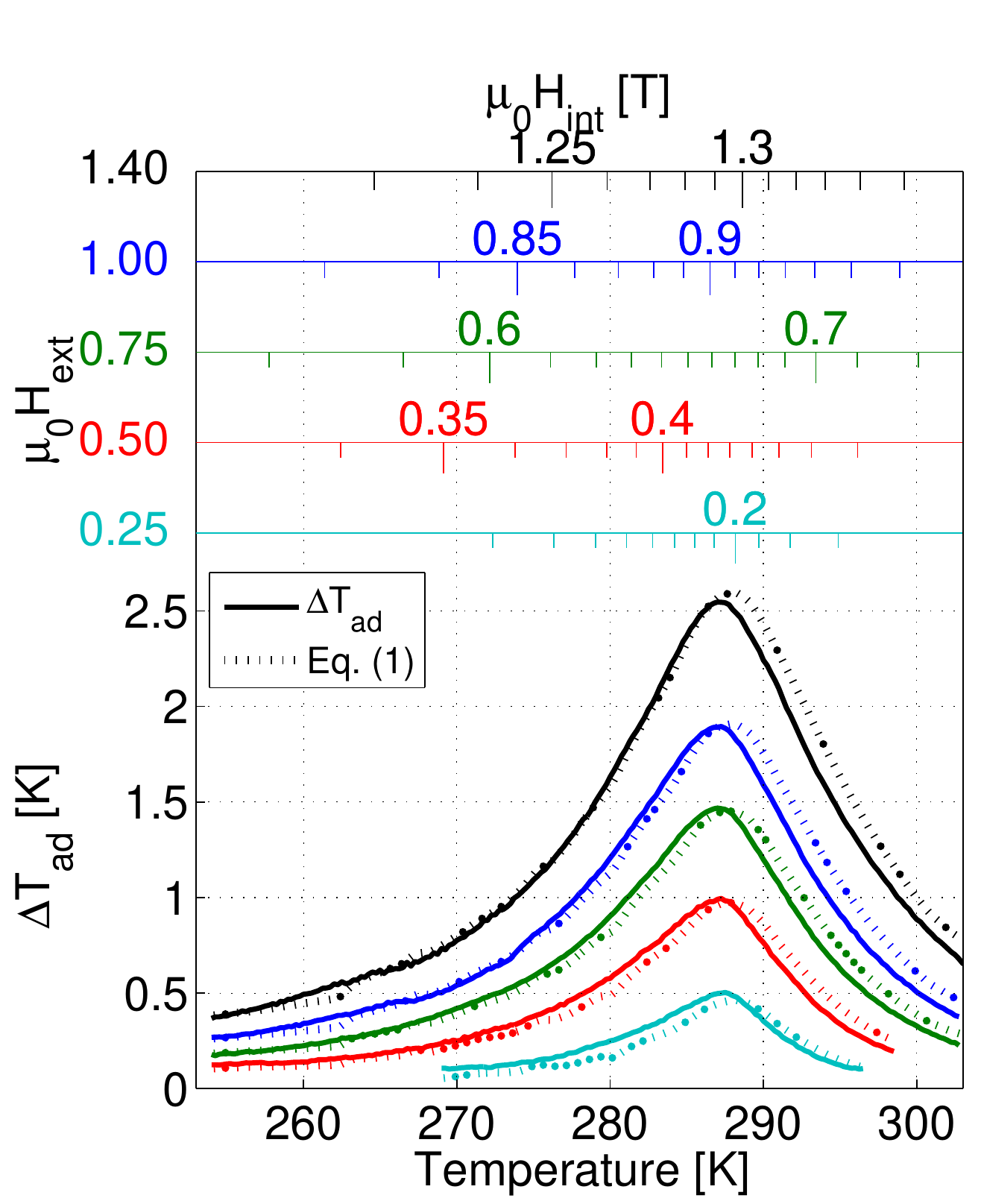}}\\
\subfigure[Adiabatic temperature change of Sample 3.]{\includegraphics[width=0.47\textwidth]{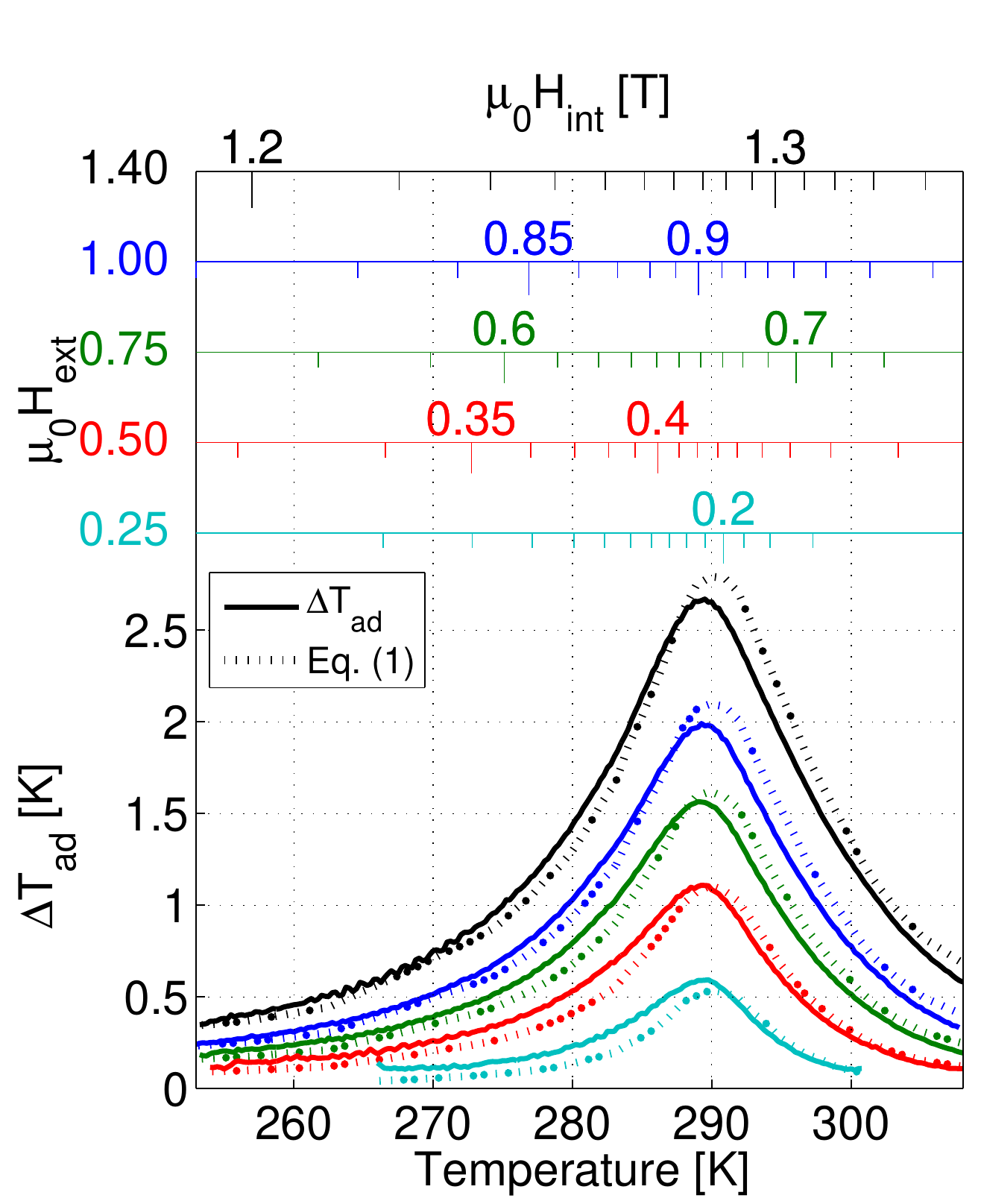}}\hspace{0.4cm}
\subfigure[Adiabatic temperature change of Gd]{\includegraphics[width=0.47\textwidth]{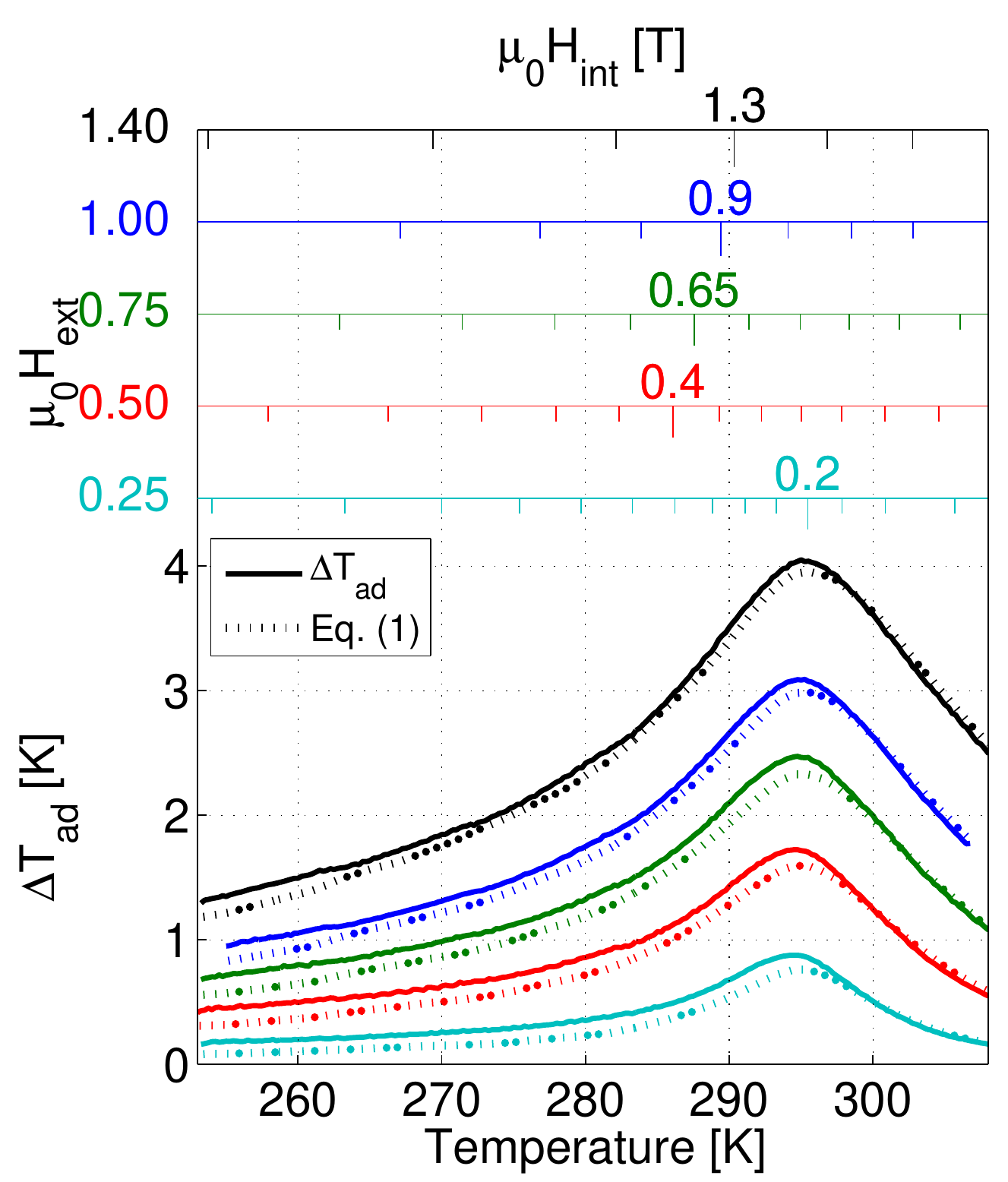}}
\caption{(Color online) The adiabatic temperature change, $\Delta{}T_\n{ad}$, as a function of temperature for a change in magnetic field from zero to the internal field given in the top x-axis. Each of the top x-axis gives the internal field for the same colored curve. The figures also show the calculated $\Delta{}T_\n{ad}$ based on Eq. (\ref{Eq_dTad}). The adiabatic temperature change increases with increasing external field.}\label{Fig.DeltaTad}
\end{figure*}

The measured values of $c_\mathrm{p}$, corrected for demagnetization and binned in 0.25 K intervals, are shown in Figure \ref{Fig.cp}. The 95\% confidence interval resulting from the binning of the data is of the order of the width of the plotted lines.

The measured values of $\Delta{}T_\mathrm{ad}$, corrected for demagnetization and binned in 0.25 K intervals, are shown in Figure \ref{Fig.DeltaTad}. Again the 95\% confidence interval resulting from the binning of the data is of the order of the width of the plotted lines. The adiabatic temperature has also been calculated using Eq. (\ref{Eq_dTad}) using the measured magnetization and specific heat capacities. The results of these calculation are also shown in Figure \ref{Fig.DeltaTad}. In general a very good agreement between the calculated and measured values of $\Delta{}T_\n{ad}$ is seen.

The adiabatic temperature change upon removal of the magnetic field has also been measured, i.e. $\Delta{}T_\n{ad,field\;off}$, although these are not shown in Figure \ref{Fig.DeltaTad}. A requisite for the reversibility of the MCE is that $\Delta{}T_\n{ad,field\;on}(T) = -\Delta{}T_\n{ad,field\;off}(T+\Delta{}T_\n{ad,field\;on}(T))$. This relation has been found to be true for all the measured data, and thus the MCE is reversible.

In both Figs. \ref{Fig.cp} and \ref{Fig.DeltaTad} the top x-axes show the internal magnetic field of the different samples, i.e. the external field corrected for demagnetization. For a given external field the internal field is a function of temperature, as the magnetization changes with temperature. Thus, here one can directly see how important it is to correct for demagnetization.

The position of the peak of both $\Delta{}s$, $c_\n{p}$ and $\Delta{}T_\n{ad}$ changes between the three different samples of \LaFeCoSi{}. Thus it is clearly seen that the peak position is tuneable.

It is known from literature that \LaFeCoSi{} may contain an impurity phase of $\alpha$-Fe. However, this will in general not greatly affect the magnetocaloric properties of \LaFeCoSi{}. This is because the specific heat capacity of $\alpha$-Fe is $c_\n{p} \sim{} 450$ J kg$^{-1}$ K$^{-1}$ which is close to observed heat capacity of \LaFeCoSi{}. The density for the $\alpha$-Fe phase is 7.87 kg m$^{-3}$ which is also close to that of the \LaFeCoSi{}.  For Sample 1 of \LaFeCoSi{} X-ray diffraction has been performed and the results were reported in Ref. \cite{Hansen_2009}. Here a 5\% $\alpha$-Fe content was identified and the structure of the \LaFeCoSi{} was found to be of NaZn$_{13}$-type. As the remaining samples are manufactured in an identical manner to Sample 1 these are presumed to have the same low content of $\alpha$-Fe as observed in similar series of \LaFeCoSi{} materials \cite{Katter_2009}.

\begin{figure}[!t]
  \centering
   \includegraphics[width=1\columnwidth]{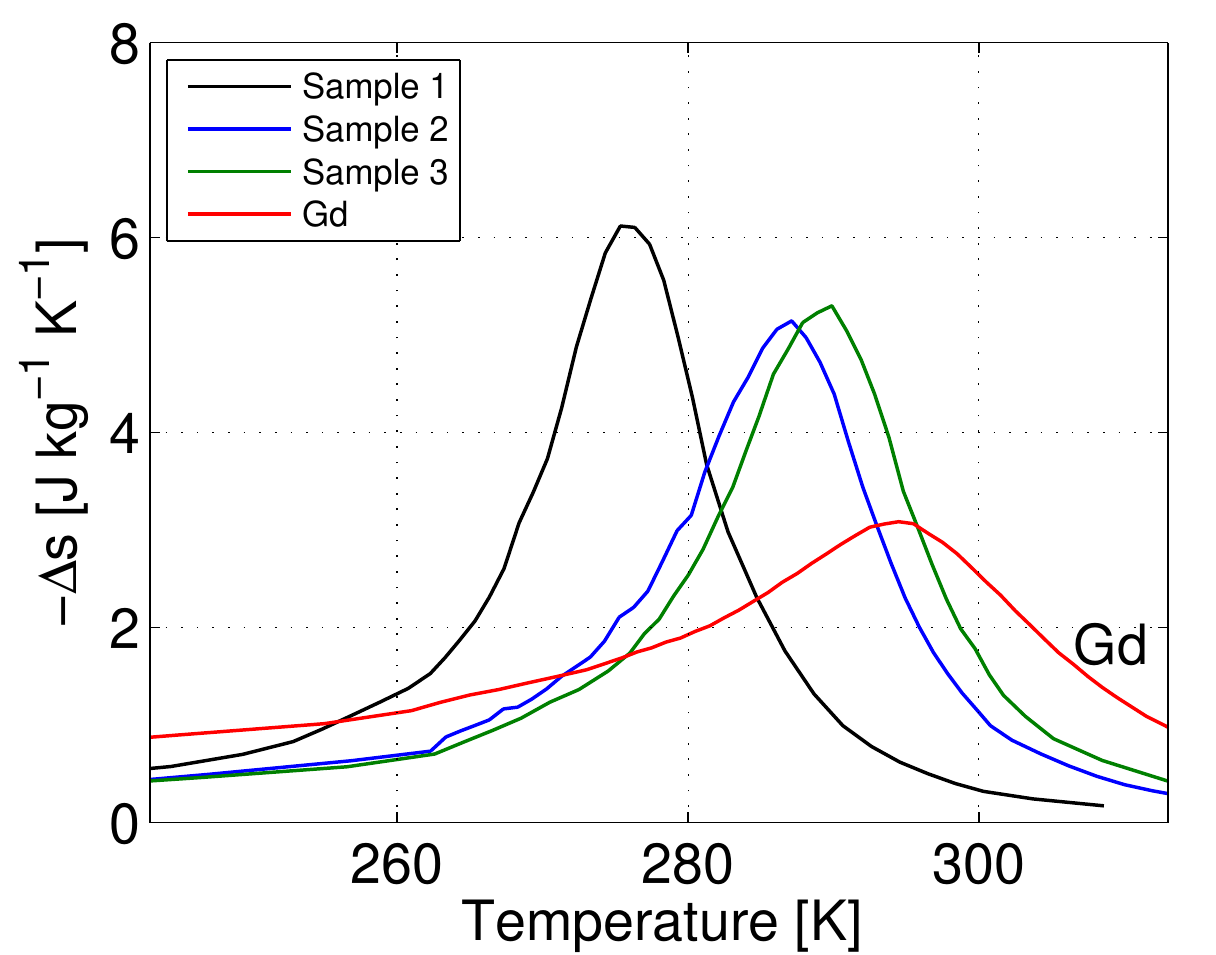}
      \caption{(Color online) The values of $\Delta{}s$ interpolated at $\mu_0{}H_\n{int}= 1$ T as a function of temperature for Gd and the three different samples of \LaFeCoSi{}. The peak temperature increases with increasing sample number for the \LaFeCoSi{} samples.}
      \label{Fig.Deltas_1T}
\end{figure}

\begin{figure}[!t]
  \centering
   \includegraphics[width=1\columnwidth]{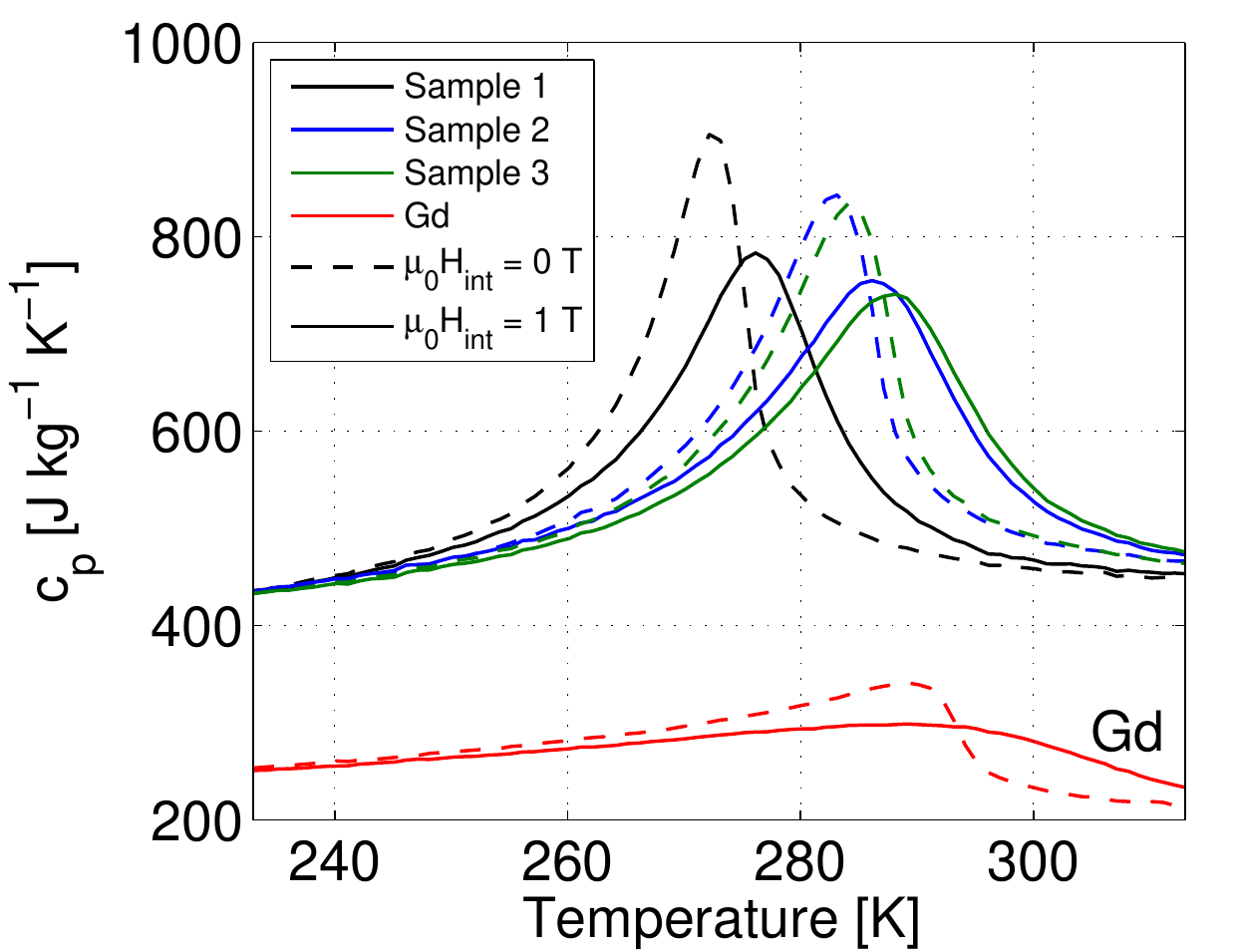}
      \caption{(Color online) The values of $c_\n{p}$ interpolated at $\mu_0{}H_\n{int}= 1$ T as a function of temperature for Gd and the three different samples of \LaFeCoSi{}. The peak temperature increases with increasing sample number for the \LaFeCoSi{} samples.}
      \label{Fig.cp_1T}
\end{figure}

\begin{figure}[!t]
  \centering
   \includegraphics[width=1\columnwidth]{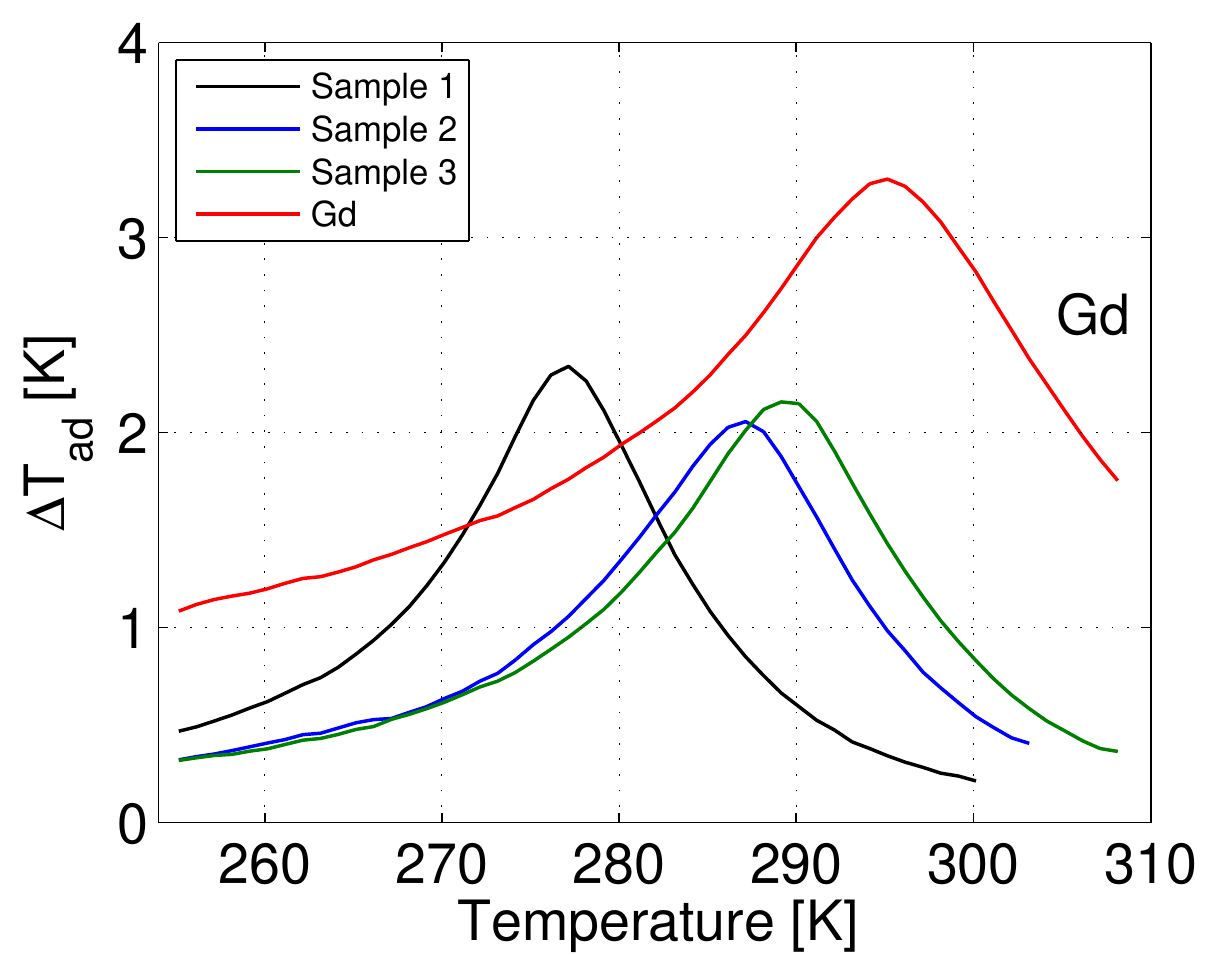}
      \caption{(Color online) The values of $\Delta{}T_\n{ad}$ interpolated at $\mu_0{}H_\n{int}= 0-1$ T as a function of temperature for Gd and the three different samples of \LaFeCoSi{}. The peak temperature increases with increasing sample number for the \LaFeCoSi{} samples.}
      \label{Fig.DeltaTad_1T}
\end{figure}

In order to compare the \LaFeCoSi{} and commercial grade Gd, each of the magnetocaloric properties has been interpolated for an internal field of 1 T. The results are shown in Figs. \ref{Fig.Deltas_1T}, \ref{Fig.cp_1T} and \ref{Fig.DeltaTad_1T} as well as given in Table \ref{Table.LaFeCoSi_and_Gd_properties}.

\begin{table*}[!t]
\begin{center}
\caption{The peak temperature and corresponding thermodynamic values for $\Delta{}s$, $c_\n{p}$ and $\Delta{}T_\n{ad}$ of \LaFeCoSi{} and Gd. The value of $\Delta{}T_{95}$ gives the band around $T_\n{peak}$ in which the value of above 95\% of the peak value. Note that the peak in $c_\mathrm{p}$ for Gd is very broad, making $T_\mathrm{peak}$ hard to determine accurately.}\label{Table.LaFeCoSi_and_Gd_properties}
\begin{tabular}{l|l|ccccr}
\multicolumn{2}{c}{Variable} & \multicolumn{3}{c}{LaFe$_{13-x-y}$Co$_x$Si$_y$} & Gd \\
\multicolumn{2}{c}{} & $\left( \begin{array}{ll} x=0.86\\ y=1.08 \end{array} \right)$ & $\left( \begin{array}{ll} x=0.94\\ y=1.01 \end{array} \right)$ & $\left( \begin{array}{ll} x=0.97\\ y=1.07 \end{array} \right)$ & $\begin{array}{cc} \mathrm{commercial}\\ \mathrm{grade} \end{array} $ &
Unit\\
\multicolumn{2}{c}{} & Sample 1 & Sample 2 & Sample 3 \\ \hline
\multirow{4}{*}{$\Delta{}s(1\; \n{T})$} & value & 6.1   & 5.1   & 5.3   & 3.1   & [J kg$^{-1}$ K$^{-1}$]\\
                                         & $T_\n{peak}$          & 275.8 & 287.1 & 289.8 & 294.8 & [K]\\
                                        & $\Delta{}T_{95}$     & -1.5/+1.9 & -1.8/+1.4 & -2.2/+1.2 & -3.4/+2.0 & [K]\\&\\
%
\multirow{2}{*}{$c_\n{p}(0\; \n{T})$}     & value &  910   & 840   & 835   & 340   & [J kg$^{-1}$ K$^{-1}$]\\
                                         & $T_\n{peak}$            & 272.6 & 283.0 & 284.3 & 289.2 & [K]\\
                                           & $\Delta{}T_{95}$     & -1.7/+1.3 & -2.5/+2.0 & -2.4/+2.0 & -6.7/+3.0 & [K]\\&\\
\multirow{2}{*}{$c_\n{p}(1\; \n{T})$}  & value & 785   & 755   & 740   & 300   & [J kg$^{-1}$ K$^{-1}$]\\
                                       & $T_\n{peak}$  & 276.1 & 286.1 & 288.2 & 289.2 & [K]\\
                                       & $\Delta{}T_{95}$     & -2.7/+2.7 & -3.5/+3.6 & -4.3/+3.1 & -19.0/+10.0 & [K]\\&\\
\multirow{2}{*}{$\Delta{}T_\n{ad}(1\; \n{T})$} & value   & 2.3   & 2.1   & 2.2   & 3.3   & [K]\\
                                                & $T_\n{peak}$   & 277.1 & 287.1 & 289.6 & 295.1 & [K]\\
                                                & $\Delta{}T_{95}$     & -1.5/+1.4 & -1.8/+1.6 & -2.1/+1.6 & -2.6/+2.6 & [K]
\end{tabular}
\end{center}
\end{table*}

From the figures it is seen that although the specific entropy change of Gd is significantly lower than for the \LaFeCoSi{} materials the adiabatic temperature change is larger over a broad interval. This is because the specific heat capacity of Gd is significantly lower than those of the \LaFeCoSi{} materials. From these facts it is clear that the magnetocaloric properties and its application potential with regard to magnetic refrigeration cannot be judged from the specific entropy change alone. Also, it is seen that if only the adiabatic temperature change is considered the Curie temperature of a \LaFeCoSi{} material must be below around 280 K for the material to outperform Gd.

\begin{figure}[!t]
  \centering
   \includegraphics[width=1\columnwidth]{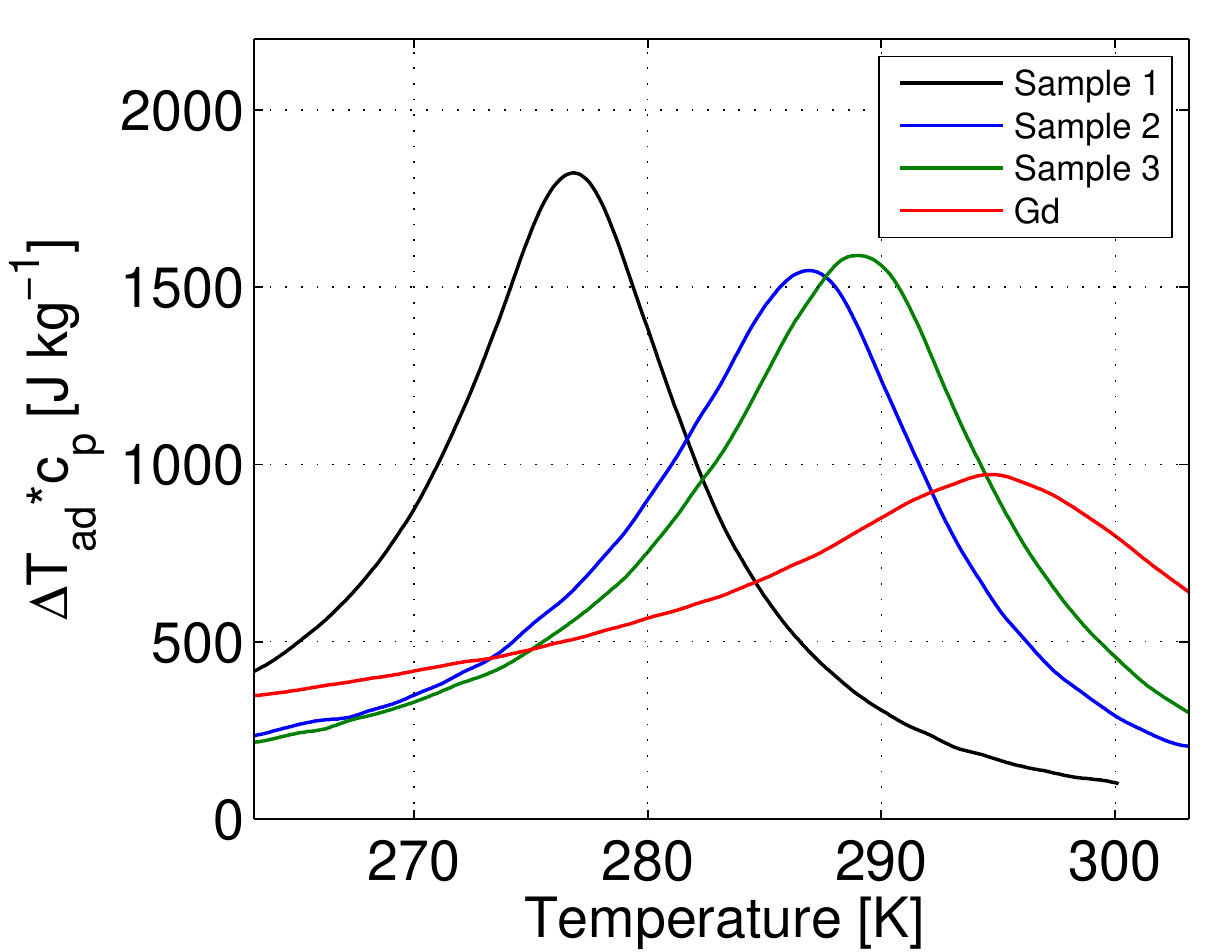}
      \caption{(Color online) The heat generated in one magnetocaloric cycle, $Q = \Delta{}T_\n{ad}c_\n{p}$, of Gd and \LaFeCoSi{} at $\mu_0{}H_\n{int}= 1$ T, based on the values shown in Figs. \ref{Fig.cp_1T} and \ref{Fig.DeltaTad_1T}.}
      \label{Fig.Heat}
\end{figure}

One can consider the heat generated per cycle of the magnetocaloric effect. This is $Q = \Delta{}T_\n{ad}c_\n{p}$ and is shown in Fig. \ref{Fig.Heat}. From this figure one can see that the \LaFeCoSi{} generates a larger heat over a large temperature interval than Gd. The cooling capacity of a regenerator is proportional to the generated heat, but the temperature span obtainable depends on the adiabatic temperature change and thus Gd will still be able to generate a larger no load temperature span than a single \LaFeCoSi{} material.

\begin{figure}[!t]
  \centering
   \includegraphics[width=1\columnwidth]{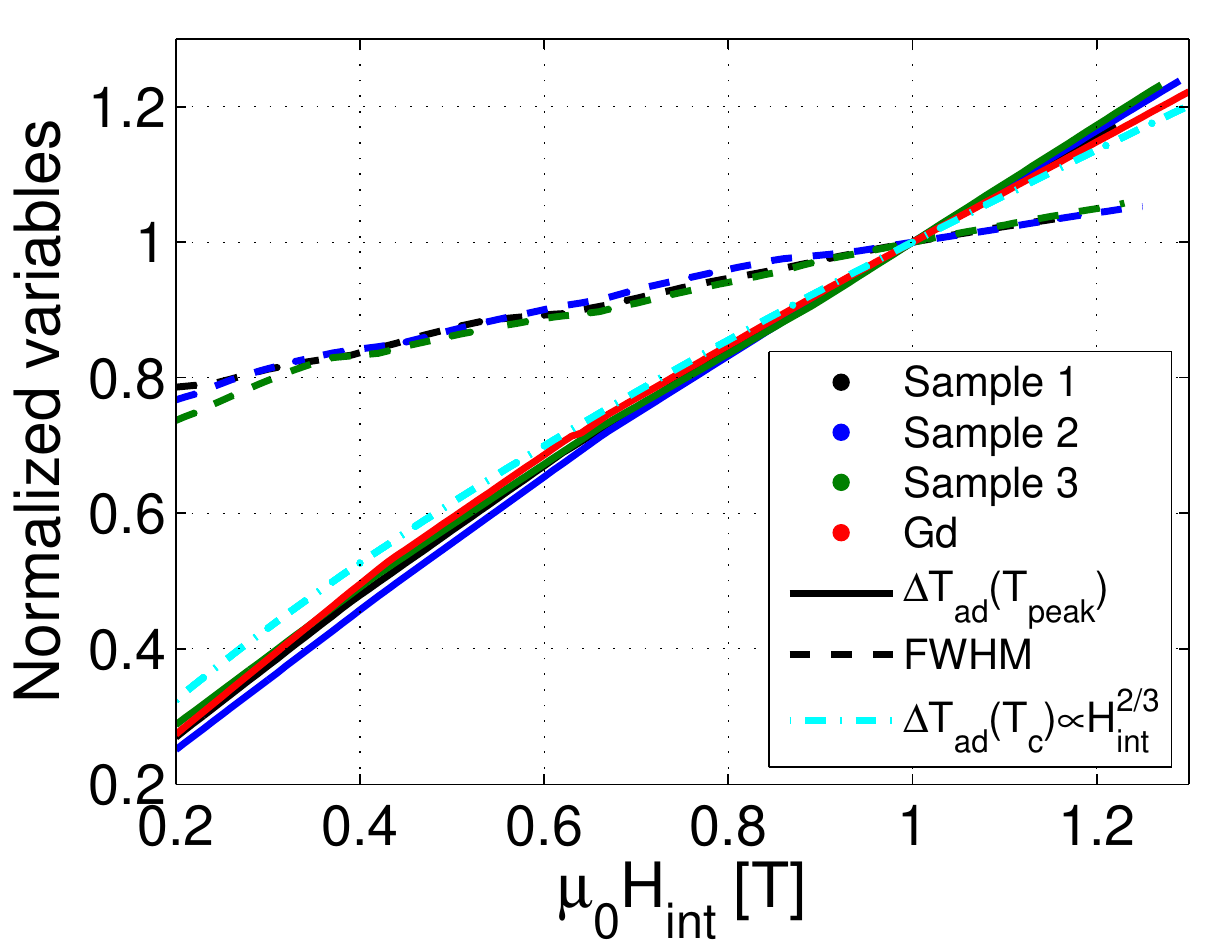}
      \caption{(Color online) The normalized peak value of $\Delta{}T_\n{ad}$ and normalized Full Width Half Maximum (FWHM) as a function of internal magnetic field for Gd and \LaFeCoSi{}, except for the Gd FWHM curve which could not be calculated because of a too small temperature range. The scaling of the adiabatic temperature change at the Curie temperature for a theoretical general second order magnetocaloric material, which scales with the magnetic field to the power of $2/3$, is also shown. The scaling of the materials are seen to be almost identical.}
      \label{Fig.Normalized_FWHM_and_DeltaTad_and_DeltaTad_area}
\end{figure}

In Figure \ref{Fig.Normalized_FWHM_and_DeltaTad_and_DeltaTad_area} the normalized interpolated value of $\Delta{}T_\n{ad}(T_\n{peak})$ and the interpolated normalized full width half maximum (FWHM) are shown as functions of internal magnetic field for Gd and \LaFeCoSi{}, except for the FWHM curve for Gd, as the temperature span is not large enough to calculate this. Here one can see that $\Delta{}T_\n{ad}(T_\n{peak})$ scales similarly for Gd and \LaFeCoSi{}. All three different \LaFeCoSi{} materials display identical behavior, as could be expected. The adiabatic temperature change of a theoretical general second order magnetocaloric material scales with the magnetic field to the power of $2/3$ \cite{Oesterreicher_1984}. This relation is also shown in Figure \ref{Fig.Normalized_FWHM_and_DeltaTad_and_DeltaTad_area}. It is seen that the scaling of both Gd and \LaFeCoSi{} are very similar, both with a scaling close to the theoretical value.

The uncertainty of the measurements performed is not trivial to estimate. For e.g. the adiabatic temperature change the uncertainty of the thermocouple temperature measurement is of the order of 0.1 K, while the measurement uncertainty for the Hall probe (AlphaLab Inc, Model: DCM) is $\pm 2$\%. Systematic errors can, however, always be present. As previously mentioned the 95\% confidence interval resulting from the binning of the data is of the order of the width of the plotted lines in Fig. \ref{Fig.DeltaTad_1T}. Thus the uncertainty of the individual measurement is small. As a measure of the uncertainty on the determination of the peak temperatures the band on which the parameters $\Delta{}s$, $c_\n{p}$ and $\Delta{}T_\n{ad}$ are within 95\% of their peak values, $\Delta{}T_{95}$, has been computed and reported in Table \ref{Table.LaFeCoSi_and_Gd_properties}. These values give the width of the peak and can be seen to change depending on the type of variable and material measured. Especially the 95\% values of $c_\n{p}(1\;\n{T})$ are seen to be quite large and thus here the uncertainty in the peak temperature is larger than for other types of variables and materials.


\section{Conclusion}
The magnetocaloric properties of Gd and three sample of \LaFeCoSi{} with different chemical composition have been measured directly. The measurements were corrected for demagnetization, allowing the data to be directly compared. In an internal field of 1 T the specific entropy change was 6.2, 5.1 and 5.0 J/kg K, the specific heat capacity was 910, 840 and 835 J/kg K and the adiabatic temperature change was 2.3, 2.1 and 2.1 K for the three \LaFeCoSi{} samples respectively. The peak temperature changes of the order of 1 K depending on the property measured, but are around 276, 286 and 288 K for the three samples respectively. The corresponding values for all properties for Gd are 3.1 J/kg K, 340 J/kg K, 3.3 K and a peak temperature of 295 K. Thus, \LaFeCoSi{} has a large enough magnetocaloric effect for practical application in magnetic refrigeration. Finally, an excellent agreement between the calculated adiabatic temperature change using Eq. (\ref{Eq_dTad}) and the measured adiabatic temperature change was seen.

\section*{Acknowledgements}
The authors would to thank J. Geyti for his technical help, and Dr. N. Pryds and Dr. A. Smith for fruitful discussions. The authors would like to acknowledge the support of the Programme Commission on Energy and Environment (EnMi) (Contract No. 2104-06-0032) which is part of the Danish Council for Strategic Research.

\end{document}